\documentclass[biber, 12pt]{article}
\usepackage[utf8]{inputenc}
\usepackage{authblk}
\usepackage{setspace}
\usepackage[margin=1.25in]{geometry}
\usepackage{graphicx}
\usepackage{subcaption}
\usepackage{amsmath}
\usepackage{lineno}
\usepackage{lipsum}

\usepackage{amsfonts}

\newcommand{\bec}{\begin{center}}
	\newcommand{\eec}{\end{center}}
\newcommand{\beq}{\begin{equation}}
	\newcommand{\eeq}{\end{equation}}
\newcommand{\bea}{\begin{eqnarray}}
	\newcommand{\eea}{\end{eqnarray}}

\newcommand{\hf}{\frac{1}{2}}

\newcommand{\qtr}{\frac{1}{4}}
\newcommand{\psib}{{\overline{\psi}}}

\usepackage[style=nejm, 
citestyle=numeric-comp,
sorting=none]{biblatex}
\begin{document}
\title{Investigating Spontaneous SO(10) Symmetry Breaking in Type IIB Matrix Model}
\author[1*$\dagger$]{Arpith Kumar}
\author[1,2]{Anosh Joseph}
\author[1]{Piyush Kumar}
\affil[1]{Department of Physical Sciences, 
	Indian Institute of Science Education and Research (IISER) Mohali, Knowledge City, Sector 81, SAS Nagar, Punjab 140306, India}
\affil[2]{National Institute for Theoretical and Computational Sciences, School of Physics and Mandelstam Institute for Theoretical Physics, University of the Witwatersrand, Johannesburg, Wits 2050, South Africa}
\affil[*]{arpithk.phy@gmail.com}
\affil[$\dagger$]{Contribution to the proceedings of the XXV DAE-BRNS HEP Symposium 2022, 12-16 December 2022, IISER Mohali, India.}
\date{}
\onehalfspacing
\maketitle

\begin{abstract}

Non-perturbative formulations are essential to understand the dynamical compactification of extra dimensions in superstring theories. The type IIB (IKKT) matrix model in the large-$N$ limit is one such conjectured formulation for a ten-dimensional type IIB superstring. In this model, a smooth spacetime manifold is expected to emerge from the eigenvalues of the ten bosonic matrices. When this happens, the SO(10) symmetry in the Euclidean signature must be spontaneously broken. The Euclidean version has a severe sign problem since the Pfaffian obtained after integrating out the fermions is inherently complex. In recent years, the complex Langevin method (CLM) has successfully tackled the sign problem. We apply the CLM method to study the Euclidean version of the type IIB matrix model and investigate the possibility of spontaneous SO(10) symmetry breaking. In doing so, we encounter a singular-drift problem. To counter this, we introduce supersymmetry-preserving deformations with a Myers term. We study the spontaneous symmetry breaking in the original model at the vanishing deformation parameter limit. Our analysis indicates that the phase of the Pfaffian induces the spontaneous SO(10) symmetry breaking in the Euclidean type IIB model.
\end{abstract}

\section{Euclidean Type IIB Matrix Model}

The partition function of the Euclidean IKKT matrix model \cite{Ishibashi:1996xs}, with the action 
\beq
S_{\rm IKKT} = S_\text{b} + S_\text{f}
\eeq 
is given by 
\beq
Z = \int dX d\psi \exp(-S_{\rm IKKT}) =  \int dX ~ {\rm Pf} \mathcal{M}~ \exp (-S_{\rm b}),
\eeq
where the bosonic and fermionic actions, respectively, are
\beq
S_\text{b} = -\qtr N ~\text{tr} \left([X_\mu, X_\nu]^2 \right)~
\eeq
and
\beq
S_\text{f} = -\hf N ~\text{tr} \left(\psi_\alpha (\mathcal{C} \Gamma^\mu)_{\alpha\beta}[X_\mu,\psi_\beta] \right).
\eeq
The fermion operator $\mathcal{M}$ has the elements
\beq
\mathcal{M}_{\alpha a, \beta b} =  \frac{N}{2} \Gamma_{\alpha \beta}^\mu ~{\rm tr} \Big( X_\mu  \left[ t^a,t^b \right] \Big).
\eeq

The $N \times N$ traceless Hermitian matrices $X_\mu$ and the Majorana-Weyl spinors $\psi_\alpha$ transform as vectors and spinors, respectively, under SO(10) symmetry. We have Weyl projected representation of gamma matrices $\Gamma_{\mu}$ in 10D. The partition function, after integrating out the fermions, involves $\mathcal{M}$, which is a complex anti-symmetric matrix of size $16(N^2 - 1) \times 16(N^2 - 1)$ written in terms of the $N^2 - 1$ generators $\{t^a\}$ of SU($N$). Studies in the recent past \cite{Krauth:1998xh, Ambjorn:2000dx, Nishimura:2011xy} have highlighted the crucial role of the complex phase of the Pfaffian in the spontaneous symmetry breaking (SSB) of SO(10) symmetry. Its wild fluctuations indicate a severe {\it sign problem}, making phase-quenched approximations inaccurate. The complex Langevin method \cite{Parisi:1983mgm, Klauder:1983nn} is one of the most promising approaches but faces the obstacle of the {\it singular-drift problem} when applied to the Euclidean IKKT model. In recent inspiring studies, mass deformation has been suggested to avoid this problem \cite{Ito:2016efb, Anagnostopoulos:2017gos, Anagnostopoulos:2020xai}. In this work, we suggest supersymmetry (SUSY) preserving deformations to shift the eigenvalues of $\mathcal{M}$ away from the origin.

\section{Supersymmetry Preserving Mass Deformations}
\label{sec:susy-def}

We introduce a SUSY preserving deformation to the action of the IKKT model \cite{Bonelli:2002mb}. The deformed action is 
\beq
S = S_{\rm IKKT} + S_{\Omega},
\eeq
where
\beq
S_{\rm  \Omega} = N~ {\rm tr} \left( M^{\mu \nu} X_{\mu} X_{\nu} +i N^{\mu \nu \sigma} X_\mu \left[X_{\nu}, X_{\sigma}  \right]  +\frac{i}{8} \psib N_3 \psi \right),
\eeq
\beq
N_{3} = -\Omega\Gamma^{8}{\Gamma^{9}}^{\dagger}\Gamma^{10}, 
\eeq
\beq
N^{\mu \nu \sigma} =  \frac{\Omega}{3!} \sum_{\mu,\nu,\sigma=8}^{10}{\epsilon^{\mu \nu \sigma}}  
\eeq
and
\beq
M =  \frac{\Omega^2}{4^3} \left( \mathbb{I}_7 \oplus 3\mathbb{I}_3  \right).
\eeq
In the above, ${\epsilon}^{\mu \nu \sigma}$ is a totally anti-symmetric 3-form, $M^{\mu \nu}$ is the mass matrix and $\Omega$ is the deformation parameter. The IKKT matrix model corresponds to the $\Omega \to 0$ limit. 

\begin{figure}[h]
    \centering		
    \includegraphics[width=.48\textwidth,origin=c,angle=0]{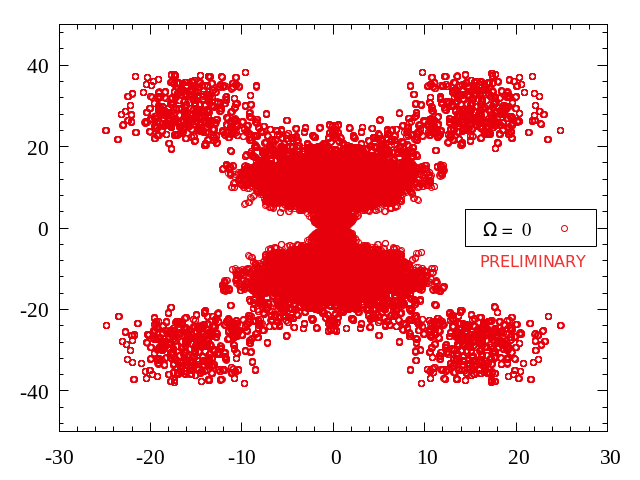}
    \includegraphics[width=.48\textwidth,origin=c,angle=0]{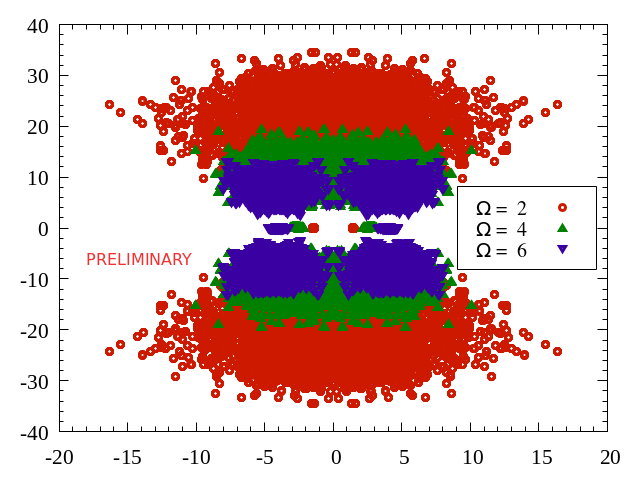}

    \includegraphics[width=.6\textwidth,origin=c,angle=0]{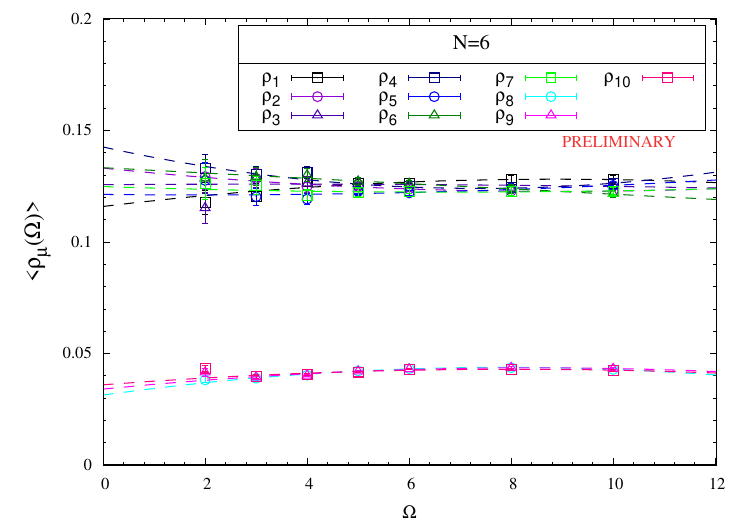}
    \caption{(Top) Scatter plots of the fermion operator eigenvalues for various $\Omega$. (Bottom) Behavior of $\rho_{\mu}$ against $\Omega$. All the simulations were done for $N = 6$. These plots are taken from Ref. \cite{Kumar:2022giw}.}	
    
    \label{fig}
\end{figure}

Our preliminary simulation results are shown in Fig. \ref{fig}. On the top plot, we see that the singular-drift problem is apparent for $\Omega = 0$, but as $\Omega$ is increased it is avoided successfully. On the bottom plot, we show the behaviour of the order parameter $\langle \rho_\mu (\Omega) \rangle$ against $\Omega$. A spontaneous breaking of SO(10) $\to$ SO(7) $\times$ SO(3) is evident even for $N = 6$. As $\Omega \to 0$, the SO(7) symmetry further breaks down into smaller subgroups, suggesting an SO$(d)$ symmetric vacuum with $d < 7$.

\section{Conclusions and Future Prospects}
\label{sec:conclusion}

We have conducted a first-principles study of the Euclidean Type IIB matrix model using the complex Langevin method. The singular drift problem was resolved using SUSY preserving deformations. Our study indicates that the Pfaffian phase triggers the spontaneous breaking of SO(10) symmetry. A comprehensive analysis of this work will appear soon elsewhere. (See Ref. \cite{Kumar:2022giw} for a recent update on this ongoing work.) We think that large-$N$ extrapolations would be necessary to understand the precise nature of the vacuum structure of this theory.

\section{Acknowledgements}
\label{sec:conclusion}

AK was partially supported by IISER Mohali and the Council of Scientific and Industrial Research (CSIR), Government of India, Research Fellowship (No. 09/ 947(0112)/2019-EMR-I). The work of AJ was supported in part by IISER Mohali and the University of the Witwatersrand. PK was partially supported by the INSPIRE Scholarship for Higher Education by the Department of Science and Technology, Government of India. We acknowledge the National Supercomputing Mission (NSM) for providing computing resources through the PARAM Smriti supercomputing system at NABI Mohali.

\printbibliography
\end{document}